\pdfoutput=1

\documentclass{PoS}

\let\subsection\undefined

\newcommand{\subsection}[1]{%
  \pagebreak[2]
  \refstepcounter{subsection}
  \addcontentsline{toc}{subsection}{
    {\protect\makebox[0.3in][r]{\thesubsection} \hspace*{3pt}#1}}
  \noindent
  \textbf{{#1}\hspace{3pt}  --}
}

\usepackage{dirtree}

\DTsetlength{0.0em}{0.1em}{0.3em}{0.5pt}{2pt}
\usepackage{siunitx}

\usepackage{graphicx}
\usepackage[font=footnotesize,labelfont=bf]{caption}
\usepackage[font=footnotesize,labelfont=bf]{subcaption}

\usepackage{wrapfig}

\graphicspath{{graphics/}}

\title{Experiences with OpenMP in tmLQCD}

\ShortTitle{Experiences with OpenMP in tmLQCD}

\author{A.~Deuzeman\\
		Albert Einstein Center for Fund. Physics, University of Bern, CH-3012 Bern, Switzerland
		E-mail: \email{albert.deuzeman@gmail.com} }
\author{K.~Jansen\\
        NIC, DESY, Zeuthen, Platanenallee 6, D-15738 Zeuthen, Germany\\
        E-mail: \email{karl.jansen@desy.de} }
\author{\speaker{B. Kostrzewa}\\
        Humboldt Universit\"at zu Berlin, Institut f\"ur Physik, Newstonstr. 15, 12489 Berlin, Germany \\
        NIC, DESY, Zeuthen, Platanenallee 6, D-15738 Zeuthen, Germany\\
        E-mail: \email{bartosz.kostrzewa@desy.de} }
\author{C.~Urbach\\
        HISKP (Theory), Rheinische Friedrich-Wilhelms Universit\"at Bonn, Germany\\
        E-mail: \email{urbach@hiskp.uni-bonn.de}}

\abstract{\includegraphics[height=3cm]{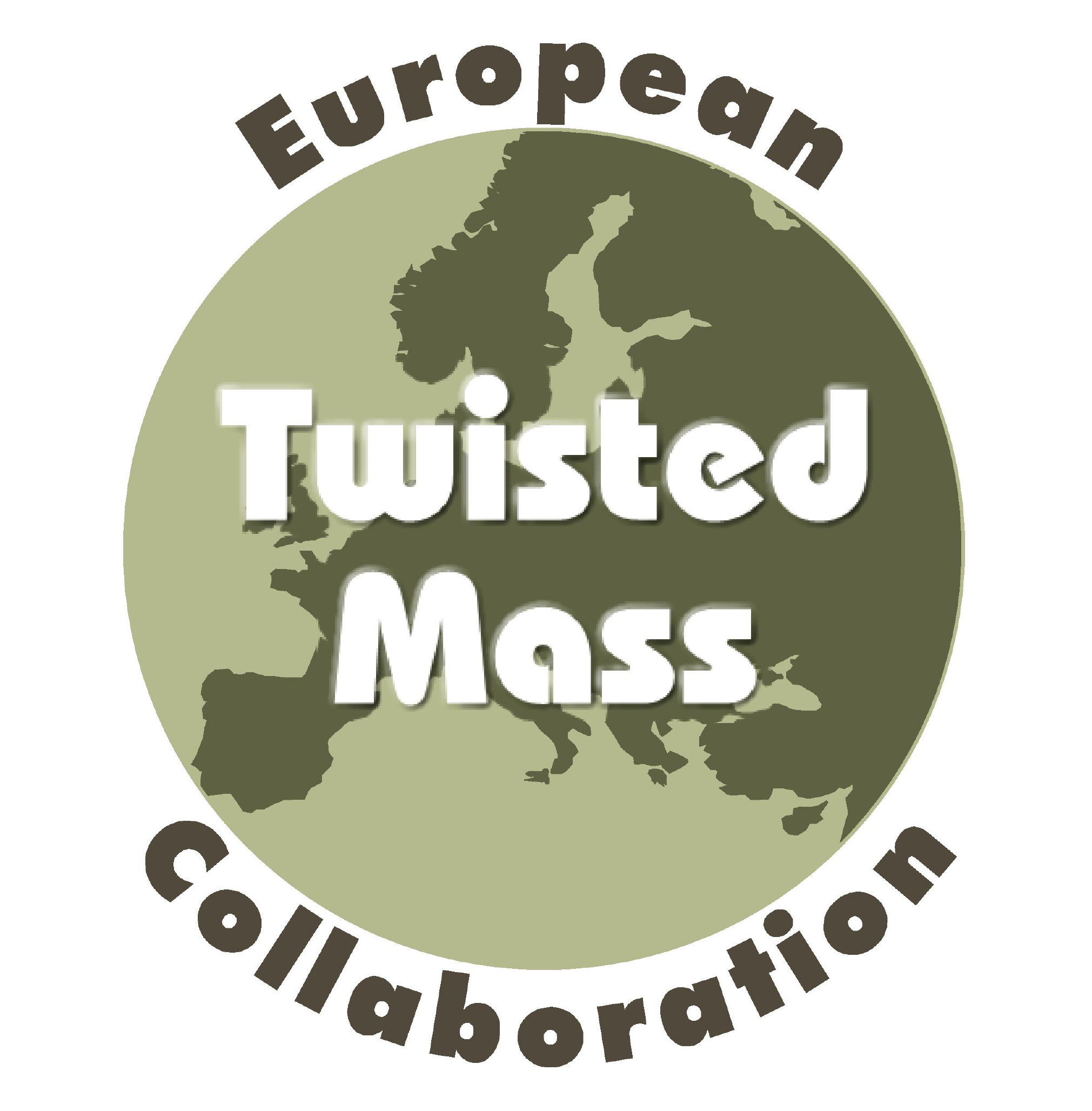}\\
An overview is given of the lessons learned from the introduction of multi-threading using OpenMP in tmLQCD.
In particular, programming style, performance measurements, cache misses, scaling, thread distribution for hybrid codes, race conditions, the overlapping of communication and computation and the measurement and reduction of certain overheads are discussed.
Performance measurements and sampling profiles are given for different implementations of the hopping matrix computational kernel.\\
\\
\footnotesize{HU-EP-13/60, DESY 13-217, SFB/CPP-13-93}}

\FullConference{31st International Symposium on Lattice Field Theory - LATTICE 2013\\
		July 29 - August 3, 2013\\
		Mainz, Germany}

\begin{document}

The tmLQCD software suite \cite{Jansen:2009tmlqcd} is a collection of programs for gauge configuration generation and Dirac matrix inversion for various types of Wilson fermions, most notably including the twisted mass term \cite{2004:rossifrezzotti1}.
For an overview of recently added features, please see \cite{2013:urbach_lat13}, presented at this conference.
In the following the source-code will be referred to frequently and the interested reader is encouraged to download or browse it on github.
\footnote{The version of the software with the overlapping kernel using MPI or SPI communication is currently found in the repository "\texttt{github.com/urbach/tmLQCD}" in the "\texttt{InterleavedNDTwistedClover}" branch.}

The hopping matrix can be found in \texttt{operator/Hopping\_Matrix.c}, which includes the files \texttt{operator/halfspinor\_body.c} and \texttt{operator/hopping\_body\_dbl.c} using 'Weyl-type' spinors with overlapping communication and computation and full four-component spinors without overlapping respectively, for a more complete explanation, see contribution \cite{2013:urbach_lat13} at this conference.
Unless mentioned otherwise, all measurements are done on 32 nodes of BlueGene/Q with 1 process per node and 64 threads per process with a node-local lattice volume of $12^4$ using the tmLQCD benchmark application.

\section{Introducing OpenMP into tmLQCD}

In the period leading up to the installation of the BlueGene/Q machine (BG/Q), JuQueen, at JSC Juelich, it was concluded that a multi-threaded version of tmLQCD was essential for good performance in the near future.
In the following, a summary of the lessons that were learned through the process of introducing OpenMP across the entire code-base will be attempted.

\subsection{Using Scoping Rules} Specifying thread-local and thread-global data in OpenMP is usually achieved through the directives \texttt{shared}, \texttt{private} and others which allow for more fine-grained control.
In C/C++, one can exploit variable scoping rules to automate the process by defining thread-local variables only inside \texttt{parallel} sections and thread-global variables only outside of them.
This way one can avoid potential bugs resulting from forgetting to update the variable lists as the code evolves.
As will be seen further below, while simplifying, this approach could however render it more difficult to introduce coarsened parallelism as an approach to reducing thread-management overhead through so-called \textit{orphaned} \cite{Tatebe:2000orphaned}  directives.
It can also be argued that the explicit listing of thread-private and thread-global variables aids in understanding, but in a language where variable scope is a central concept, this may be judged as a weak argument.

\section{Optimizing the Hopping Matrix}

\subsection{Performance Measurements} 
\begin{figure}
		\centering
		\includegraphics[width=0.85\textwidth]{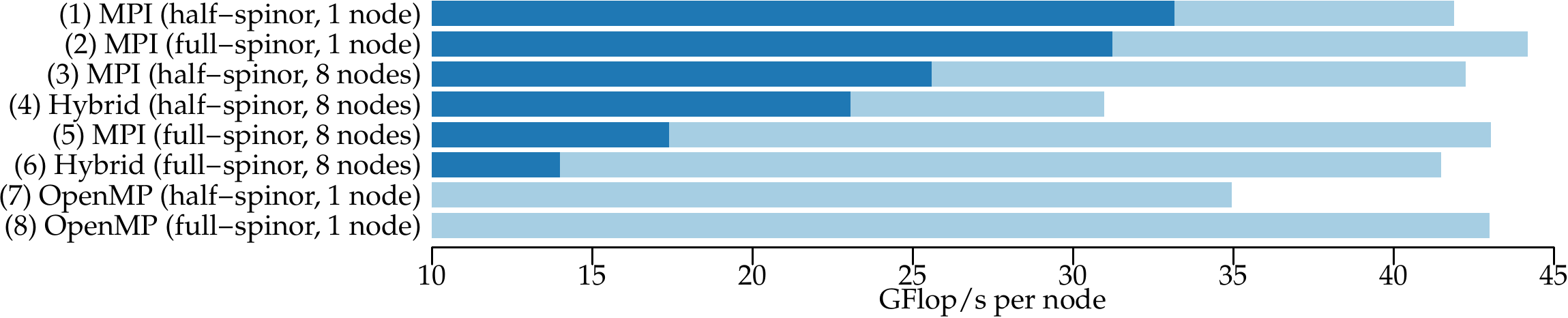}
		\caption{Performance of the tmLQCD hopping matrix on a dual-socket Intel Xeon X5560 Infiniband cluster with a lattice volume per node of $16\times8^3$. The lighter bars show performance with communication routines explicitly disabled. Hybrid refers to 2 processes per node and uses a kernel which attempts to overlap communication and computation.}
		\label{fig:Hopping:performance_pax}
\end{figure}

The hopping matrix was the first function to be multi-threaded and performance measurements were carried out on multi-core Intel\textsuperscript{\textregistered} CPUs, the results of which are shown in figure \ref{fig:Hopping:performance_pax}.
It was found that for a problem that fits into cache on a single node of this machine, the OpenMP version of both the full-spinor and the half-spinor hopping matrices outperforms the pure MPI version with a corresponding number of processes.
Once multiple nodes are required, however, the pure MPI version outperforms the hybrid OpenMP/MPI implementation on this architecture.
The authors believe that this regression originates from the loss of symmetry and additional thread synchronization required when the communication is dispatched by one of the threads.
This suspicion is supported by the measurement of the half-spinor hybrid code with communication disabled (light bar 4) but all the OpenMP structure still in place.
It is possible that optimization of MPI parameters might produce higher performance because the message size is clearly larger for the hybrid version than for the pure MPI version.
The various overheads will be discussed in detail below.

Although not studied in depth for these cases, the performance regression persists on Cray XC30 but appears to be quite mild on SuperMUC, allowing scaling to larger machine partitions through the usage of the hybrid code there.
As will be discussed in more detail below, on BG/Q the hybrid version consistently outperforms the MPI-only implementation because overlapping of communication and computation works very well on this machine.

An important conclusion that can be taken away from these results is that for hybrid parallelizations, measurements need to be carried out on each target platform taking into account different compilers, communication libraries and even the problem size as will be discussed below.

\subsection{Cache Misses} During these initial measurements it was noticed through the usage of the Intel\textsuperscript{\textregistered} VTune\textsuperscript{\texttrademark} Performance Analyzer that in the full-spinor version, a large number of cache misses and resulting thread idling occurred from lookups of neighbourhood lattice indices which are stored in two-dimensional arrays \texttt{g\_iup[x][mu]} and \texttt{g\_idn[x][mu]}.
While a direct lookup like \texttt{xp1=g\_iup[x][1]} did not pose a problem, the compiler could not optimize access to a double lookup like  \texttt{xm0p1=g\_iup[g\_idn[x][0]][1]}.

A dedicated array of indices was added, precomputed during program initialization, which contains the lattice indices in the order required by the hopping matrix and which is accessed in this way: \texttt{ix=g\_hi[16*x+n]}, where \texttt{n} is a constant integer depending on the operation in this line.
This simple optimization resulted in a performance boost of almost 20\%.

\subsection{Scaling with Thread Distribution} 
\begin{figure}
	\begin{subfigure}{0.45\textwidth}
		\centering
		\includegraphics[width=6.3cm]{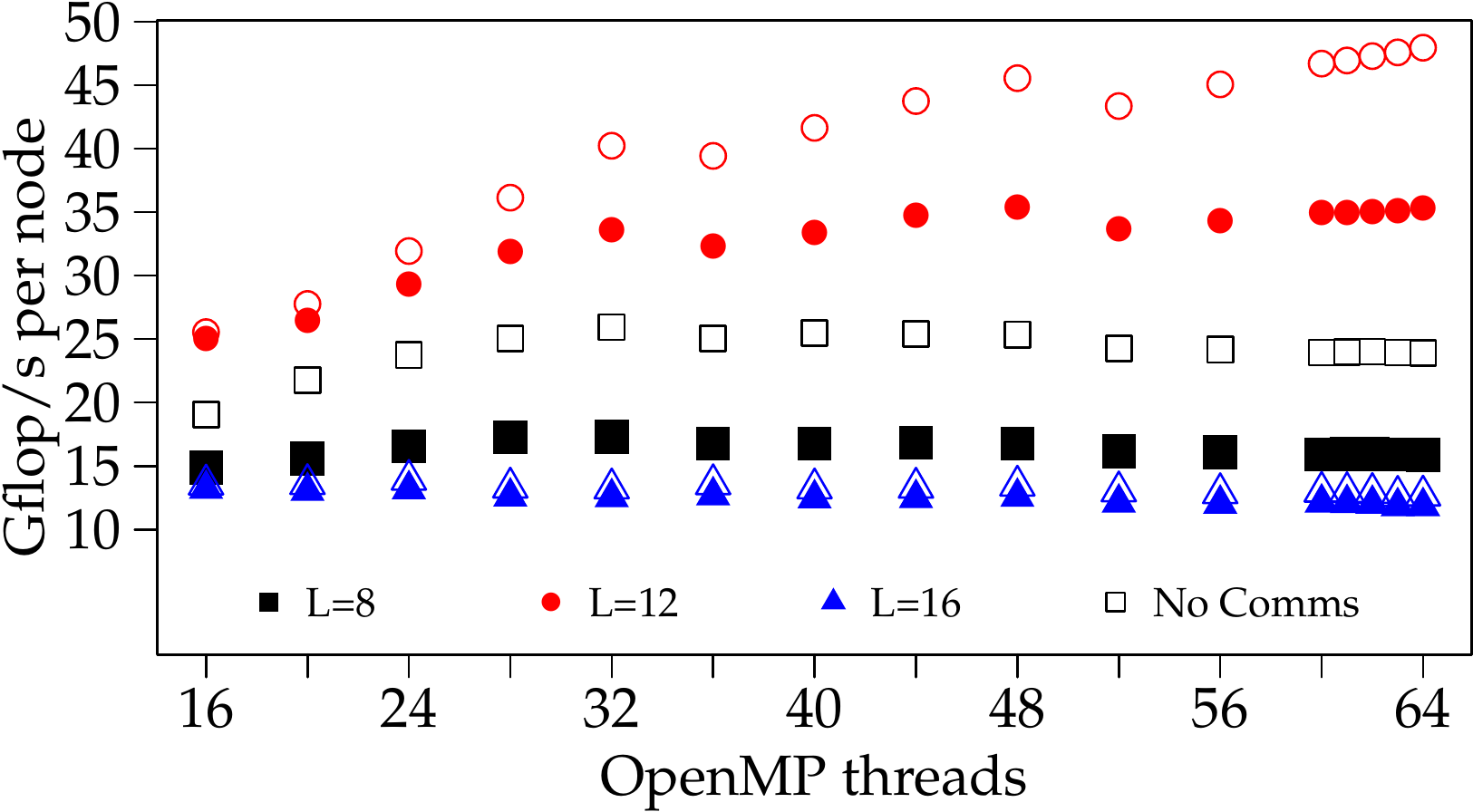}
		\caption{1 process per node and increasing numbers of threads.}
		\label{fig:HopScaling:16to64}
	\end{subfigure}
	\hspace{0.08\textwidth}
	\begin{subfigure}{0.45\textwidth}
		\centering
		\includegraphics[width=6.3cm]{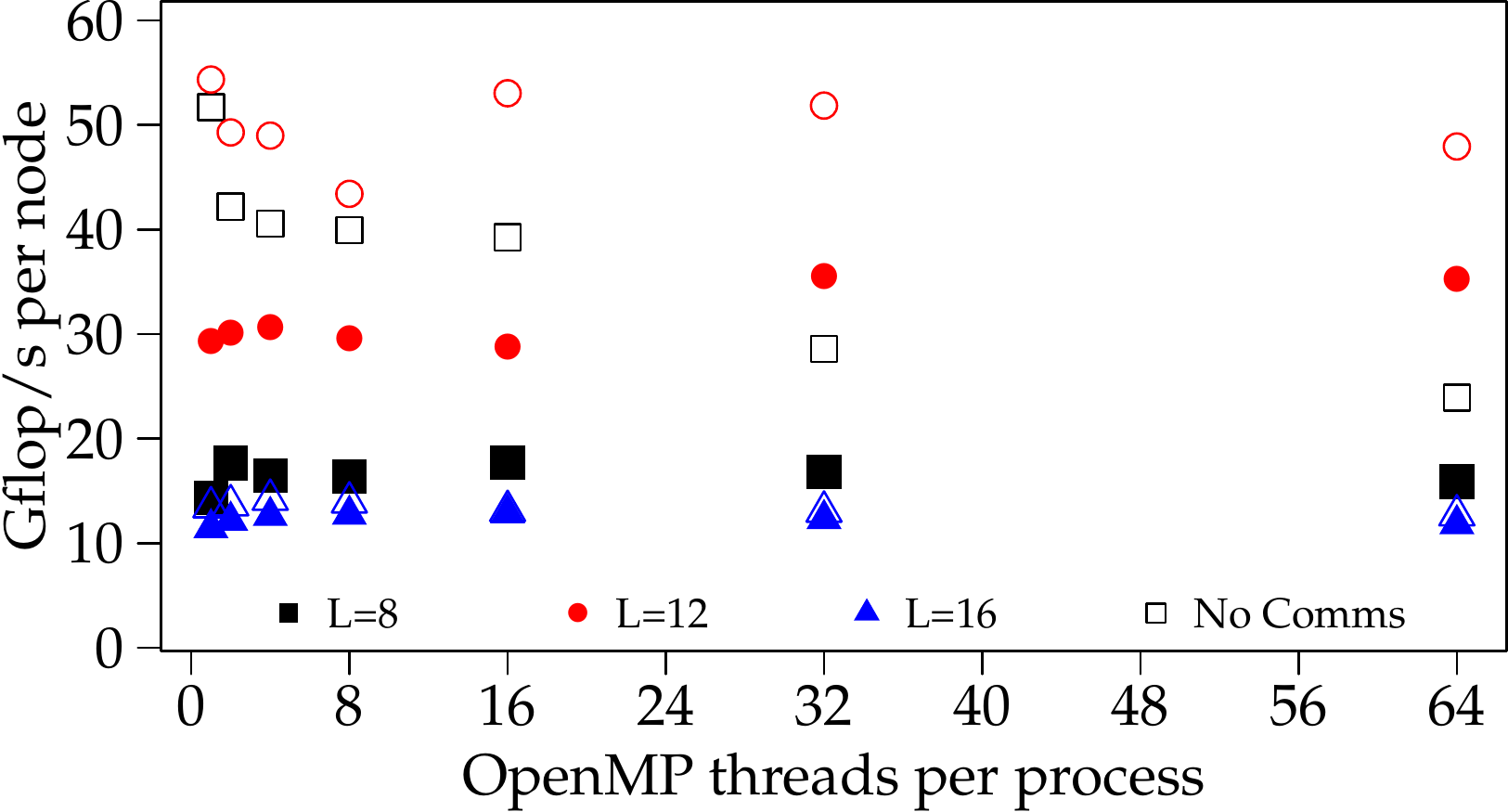}
		\caption{The number of threads times the number of processes per node is kept constant at 64.}
		\label{fig:HopScaling:const64}
	\end{subfigure}
	\caption{Performance on 32 nodes of BG/Q in GFlop/s per node  using an overlapping OpenMP/MPI hopping matrix for different node-local lattice volumes $L^4$. Empty symbols correspond to inter-process communication being disabled.}
	\label{fig:HopScaling}
\end{figure}

The performance of the hopping matrix as a function of the total number of threads and their distribution across processes was studied on BG/Q, the results are shown in figure \ref{fig:HopScaling}.
Note that the more efficient SPI communication was not used for these tests.

One conclusion which can drawn from these measurements is that for local lattice volumes which fill the L2 cache close to maximally, a configuration with 1 process per node and 64 threads is fastest.
On the other hand, when the L2 cache is not fully exploited, configurations with 4 threads per process or 32 threads in total seem to be beneficial.
Comparing in panel (b) performance with communication for $L=12$ for 16 and 32 threads per process, MPI seems to benefit from using fewer processes per node as the pure floating point performance is comparable.
On the other hand, for the $L=8$ measurement, performance without communication is severely degraded, suggesting inefficiencies in multi-threading. 
It is conceivable, therefore, that for small local lattice volumes with optimized task placement, fewer threads per process and more processes are optimal.

The reason for this situation seems clear: when there are many threads per process the OpenMP overhead is very large and this can only be mitigated by a large workload for each thread, allowing scaling (a) with the number of threads for $L=12$ which is not seen for $L=8$.
By contrast, when the local lattice volume is very small ($L=8$), it makes sense to absolutely minimize the OpenMP overhead, as shown by the significant loss of pure floating point performance as the number of threads per process is increased in panel (b).

\subsection{Race conditions and Memory Locking} Algorithms such as the hopping matrix can be efficiently implemented in a "push-style" fashion where a loop over the volume "pushes" computation results to neighbouring lattice sites.
When the work loop is shared amongst threads, as any lattice site is the neighbour of 8 other lattice sites, multiple threads can potentially attempt to write the same memory location, thus requiring some form of memory locking.
This is usually done through \texttt{critical} directives or by updating memory "atomically" using the \texttt{atomic} directive.
While the former approach is safe, it results in extreme overheads \cite{1999:bullmeasuring} which are unacceptable in a lattice QCD code.
The usage of \texttt{atomic} directives seems to have low overhead, but correctness is not guaranteed \cite[pp.380-390]{2007:bronevetsky_formal} when there are multiple writing and reading threads and the same memory location could be accessed from several \texttt{atomic} statements on different source-code lines (which would be the case for the hopping matrix, for example).
Further, the set of operations which can be carried out atomically is very limited.

\begin{wrapfigure}{r}{0.3\textwidth}
		\centering
		\includegraphics[width=0.2\textwidth]{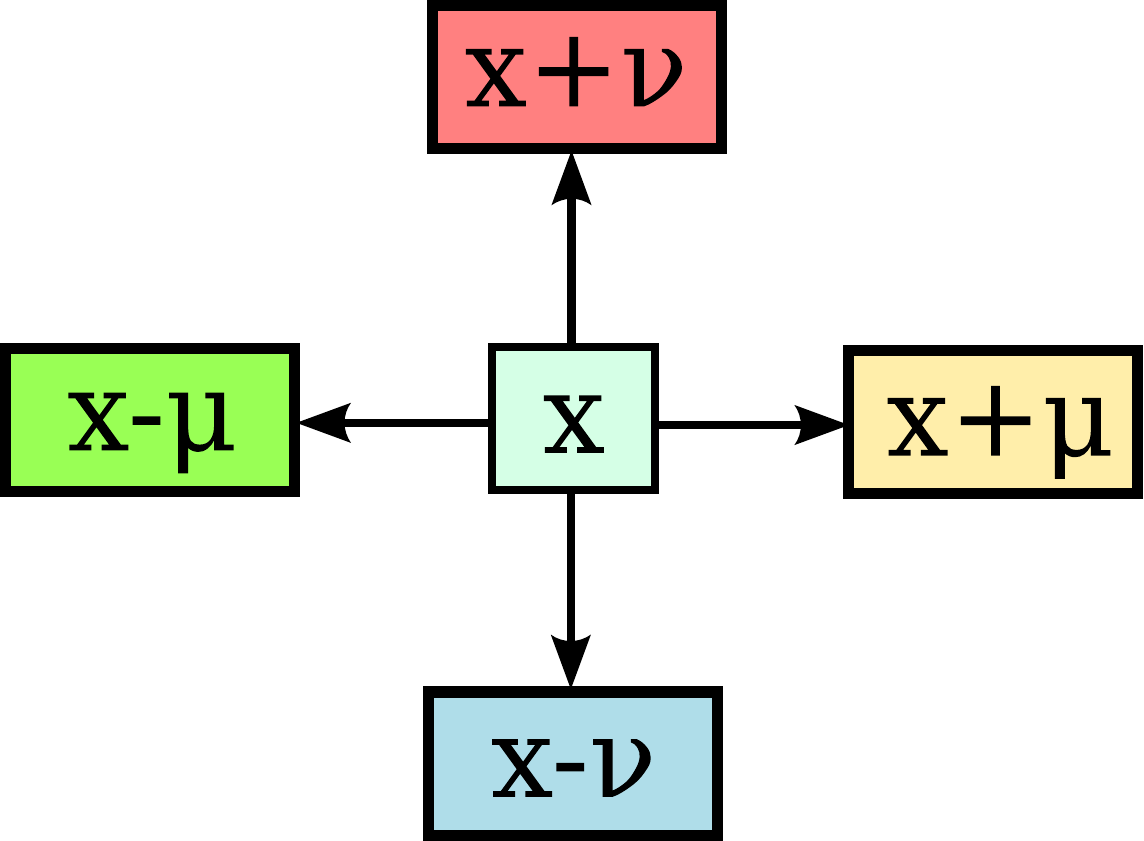}
		\caption{Pictorial representation of the concept of "halo memory".}
		\label{fig:halo}
\end{wrapfigure}

An alternative to locking is the addition of "halo memory" for \emph{each} lattice point as pictorially depicted in figure \ref{fig:halo}.
Because each lattice point now has independent neighbourhood memory, the conflict situation discussed above cannot occur.
This implementation, however, requires significantly more memory and the results for the different halos need to be accumulated in a second loop over the volume.
Since the tmLQCD half-spinor hopping matrix was already implemented in this manner even before the introduction of threads, no performance comparison has been carried out.

In the computation of the fermionic derivative during Hybrid Monte Carlo integration, tmLQCD does not use halo memory but the \texttt{atomic} directive when updating the components of the real-valued derivative field.
Very high statistics comparisons of serial and multi-threaded HMC runs have shown that there seems to be no noticeable effect of the danger discussed in \cite{2007:bronevetsky_formal}.
A dedicated study of the potential for errors from this kind of set of \texttt{atomic} statements is planned, as well as a test of halo memory for this operation.

\subsection{Overlapping Communication and Computation} The instructions for tackling this problem are usually as follows: compute the data which is to be communicated (the "surface"), start the communication using non-blocking communicators, perform the part of the computation which does not depend on communicated data (the "body"), call MPI\_Wait(all) and finally compute the remaining results.
Unfortunately it seems that in the cases that were tested by the authors (using MPI\_Isend/recv), all the communication is done during MPI\_Wait(all), no matter how much time is spent in the body computation, thereby completely defeating the purpose of attempting to overlap in the first place.

\begin{figure}
		\centering
		\includegraphics[width=0.84\textwidth]{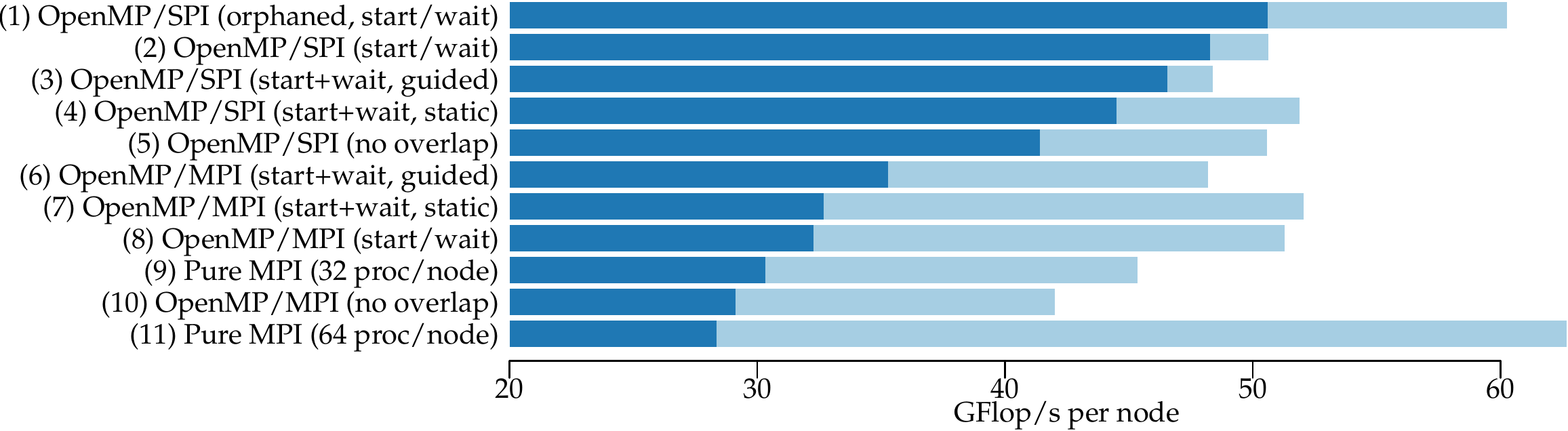}
		\caption{Performance measurements for different versions of the hopping matrix. The lighter bar shows measurements without communication. Unless noted otherwise, the hybrid kernel attempts to overlap communication and computation.}
		\label{fig:Hopping:performance}
\end{figure}

In a hybrid application, however, one could instruct one (or multiple) thread(s) to initiate the communication in a section with a \texttt{nowait} directive and then immediately call MPI\_Wait(all), thereby actually starting the communication.
When static thread scheduling is used in the body loop, this however results in a loss of performance as the other threads have to wait for the communicating thread(s) to rejoin after doing its (their) work on the body computation.
It was found that employing \texttt{guided,32} scheduling in the body computation significantly increases performance.
In addition, this implementation theoretically benefits from smaller overhead because it requires one fewer synchronization as the communicating thread(s) is (are) 'caught' by the implicit OpenMP barrier at the end of the body loop.

An overview of the structure of the "start$\rightarrow$compute$\rightarrow$wait" and "start+wait$\rightarrow$compute" implementations is shown in figure \ref{fig:Hopping:scalasca}, with the time spent in each sub-part sampled using SCALASCA \cite{Geimer:2010scalasca}.
It must be noted here that the difference in overheads between static and guided scheduling discussed above is only marginally reflected in the respective SCALASCA samplings (note: only \texttt{guided} shown here!), although the effect on performance is apparently substantial.

A summary of performance measurements on BG/Q using different implementations of an overlapping kernel is shown in figure \ref{fig:Hopping:performance}.
The hybrid code without overlapping (10) is only marginally faster than the pure MPI code (11).
A code which attempts to overlap following the 'usual' instructions is somewhat faster (8).
The implementation using only one \texttt{single} section is faster still (7) but only through the optimization of the thread scheduling is performance improved significantly (6).
Note that performance without communication drops (compare the light bars in 6 and 7), because \texttt{guided} scheduling is less efficient.
These relationships between the different hybrid implementations persist on different machines like the Cray XC30 or SuperMUC, with the caveat that the performance benefit from overlapping is only around 100 MFlop/s per core and the pure MPI version without overlapping outperforms all of them, at least on the machine partitions that were tested.

This picture is complicated by the fact that when using SPI for communication, the "naive" implementation using two \texttt{single} sections is faster than the one discussed above as shown by bars 2 and 3.
The authors assume that using SPI for Remote Direct Memory Access (RDMA) actually launches the communication without requiring many CPU cycles.
The loss of performance from \texttt{guided} scheduling in the body then causes overall performance to drop more (compare light bars in 4 and 3), than is gained by the removal of one extra synchronization (compare light bars in 4 and 2).
An RDMA implementation using one-sided MPI communicators was not attempted yet.

\section{Measuring and Reducing Overheads} 

\begin{figure}
	\begin{subfigure}{0.49\textwidth}
		\flushleft
		\includegraphics[height=6cm]{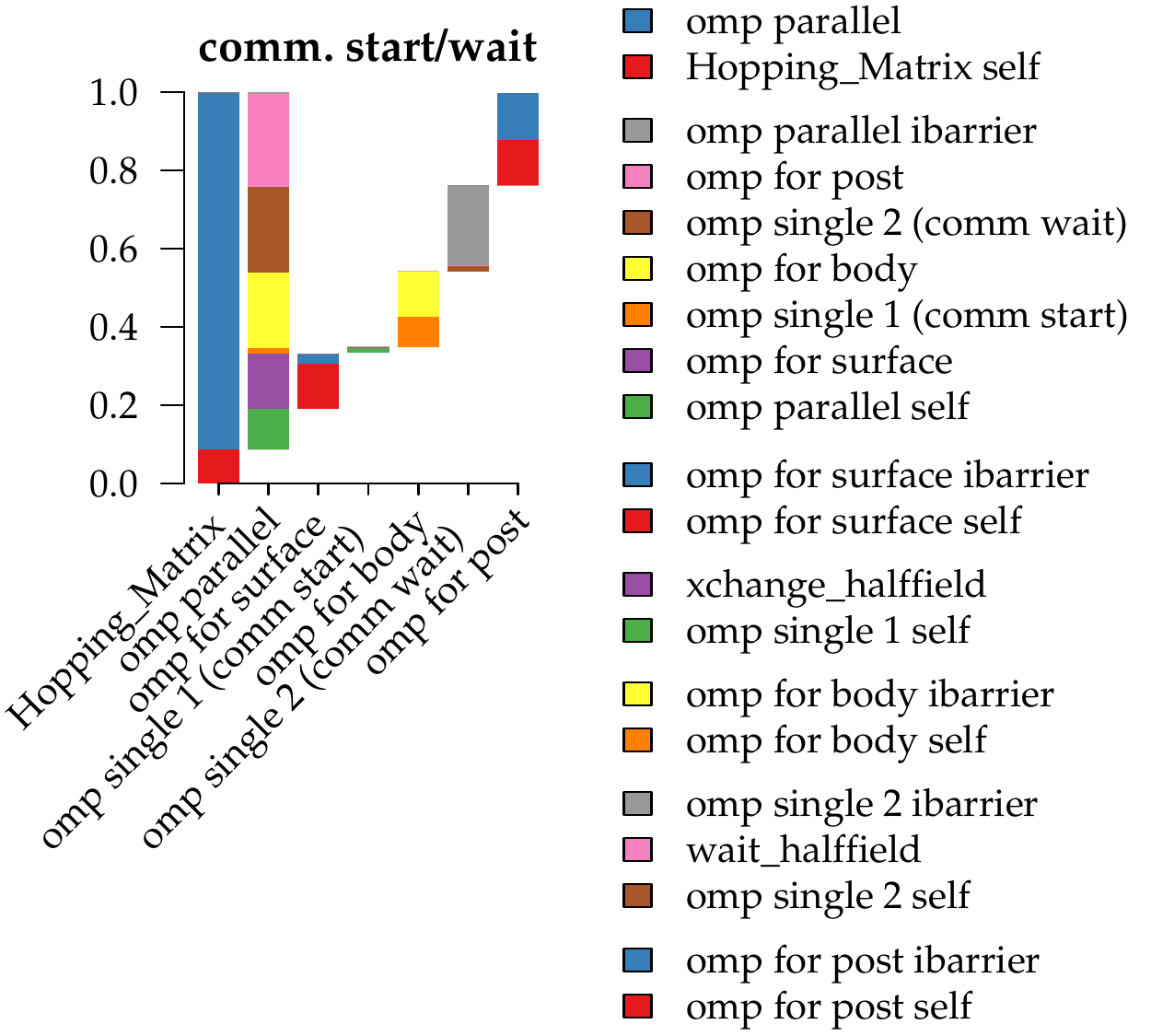}
		\caption{}
		\label{fig:Hopping:scalasca:mpi_2_single}
	\end{subfigure}
	\begin{subfigure}{0.49\textwidth}
		\flushright
		\includegraphics[height=6cm]{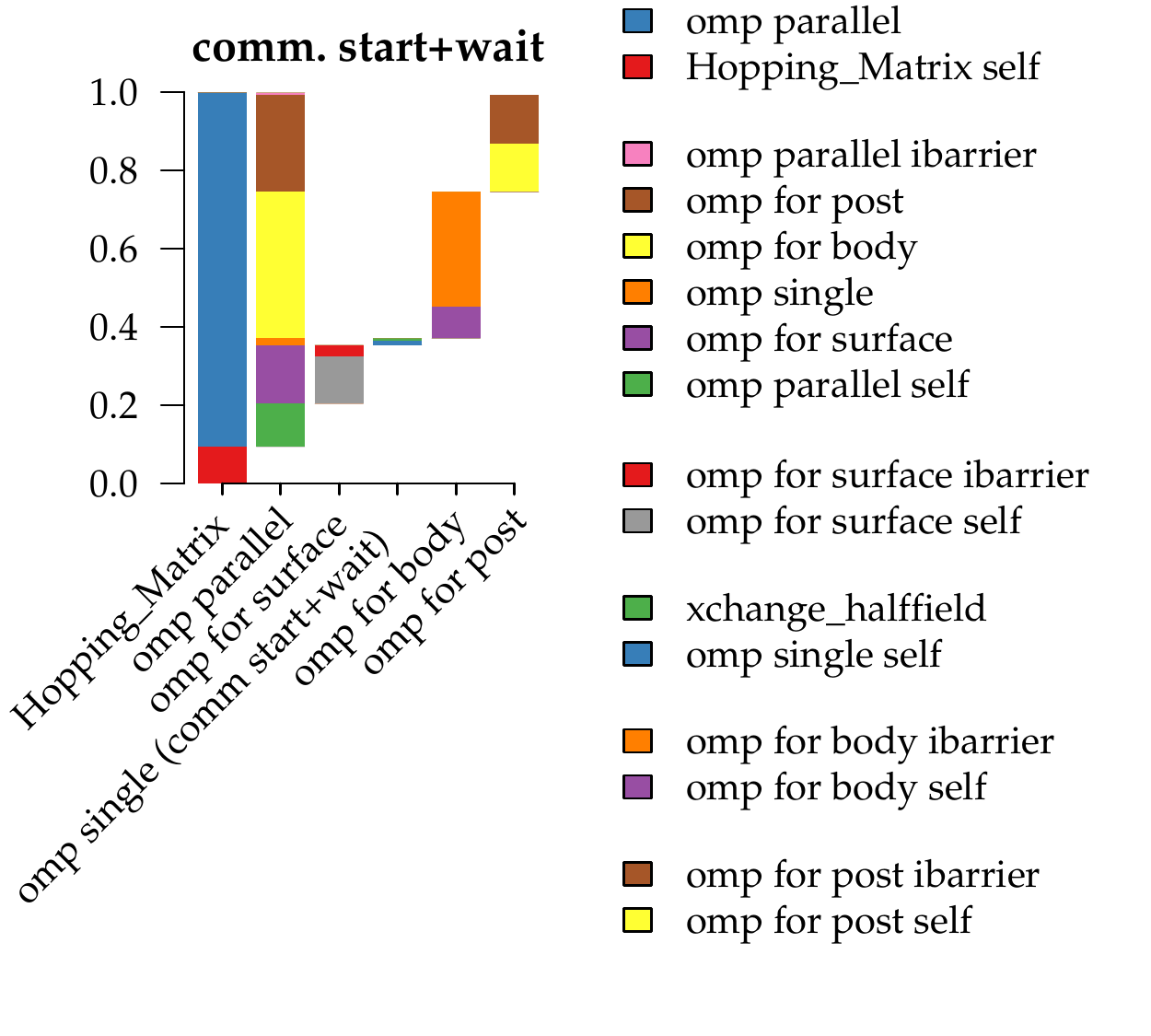}	
		\caption{}
		\label{fig:Hopping:scalasca:thread_overlap}
	\end{subfigure}
	
	\caption{SCALASCA samplings of time spent in the sub-parts of the half-spinor hopping matrix in the tmLQCD benchmark application. The legend gives the division of each bar where the qualifier "self" refers to time spent in a given part which is not spent in any of its sub-parts. Panel (a) shows the distribution in the function where non-blocking communicators are called before and MPI\_Waitall is called after the body loop. Panel (b) shows the situation when the same thread is used to start and wait for communication, with implicit synchronization at the end of the body loop.}
	\label{fig:Hopping:scalasca}
\end{figure}

The OpenMP overhead in the hopping matrix can be estimated by comparing pure floating point performance (with communication routines explicitly disabled) using 64 processes per node and the respective performance when using 1 process and 64 threads.
Comparing thus the light bars of 11 and 7 in figure \ref{fig:Hopping:performance}, one can estimate that the OpenMP overhead impacts pure floating point performance per node by around 11 GFlop/s on BG/Q.

A breakdown of the overhead can be estimated from the SCALASCA samplings shown in figure \ref{fig:Hopping:scalasca}, for instance by looking at the "self" contribution to \texttt{omp parallel} sections.
Further overheads are found in the implicit barriers marked "ibarrier" at the end of loops and \texttt{single} sections without \texttt{nowait}.
For loops, the "self" measurement represents the actual "work" done.

One surprising result from these measurements is that when the tmLQCD inverter is sampled using SCALASCA, it records a significant overhead for thread forking even though thread recycling should be taking place.
This is not seen in samplings of the tmLQCD benchmark application, which simply calls the Hopping\_Matrix function many times.
This suggests that when \texttt{parallel} sections in the same function occur repeatedly, this overhead drops out.

Still, the "self" contribution to \texttt{omp parallel} (which can be seen to reflect the thread management overhead), constitutes a significant portion of total runtime.
It can be significantly reduced through the usage of \textit{orphaned} directives, e.g. \texttt{omp pragma for} directives inside functions which are not syntactically enclosed in a \texttt{parallel} section.
Instead, threads are launched one (or several) level(s) further up, for example when a solver is called.
This has been tested for the hopping matrix benchmark, resulting in measurement 1 in figure \ref{fig:Hopping:performance}.
The underwhelming result with communication stems from a recent network performance regression on JuQueen, but the light bar indicates a significant reduction of the overhead discussed above.
Managing this kind of parallelism using OpenMP can be very demanding because thread-local and thread-global data now needs to be tracked over several function hierarchies.
Perhaps in this situation it might be beneficial to state the locality of variables using \texttt{private} and \texttt{shared} directives instead of relying on scoping rules.

\begin{wrapfigure}{r}{0.5\textwidth}
	\begin{subfigure}{0.12\textwidth}
		\flushright
		\includegraphics[clip=true,trim=0 3.8cm 0 0,height=4.4cm]{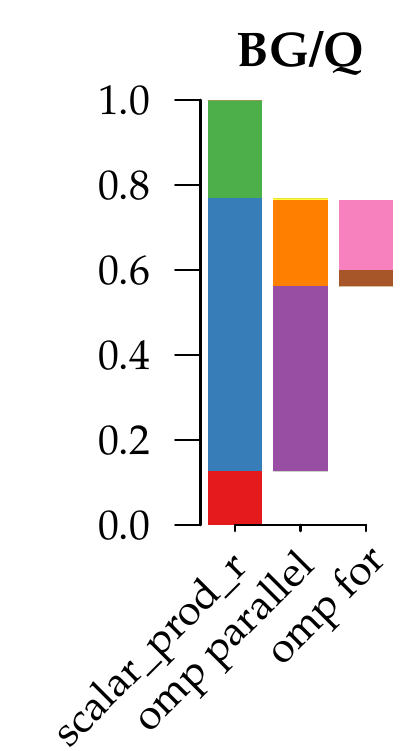}
	\end{subfigure}
	\begin{subfigure}{0.3\textwidth}
		\flushleft
		\includegraphics[clip=true,trim=0 3.8cm 0 0,height=4.4cm]{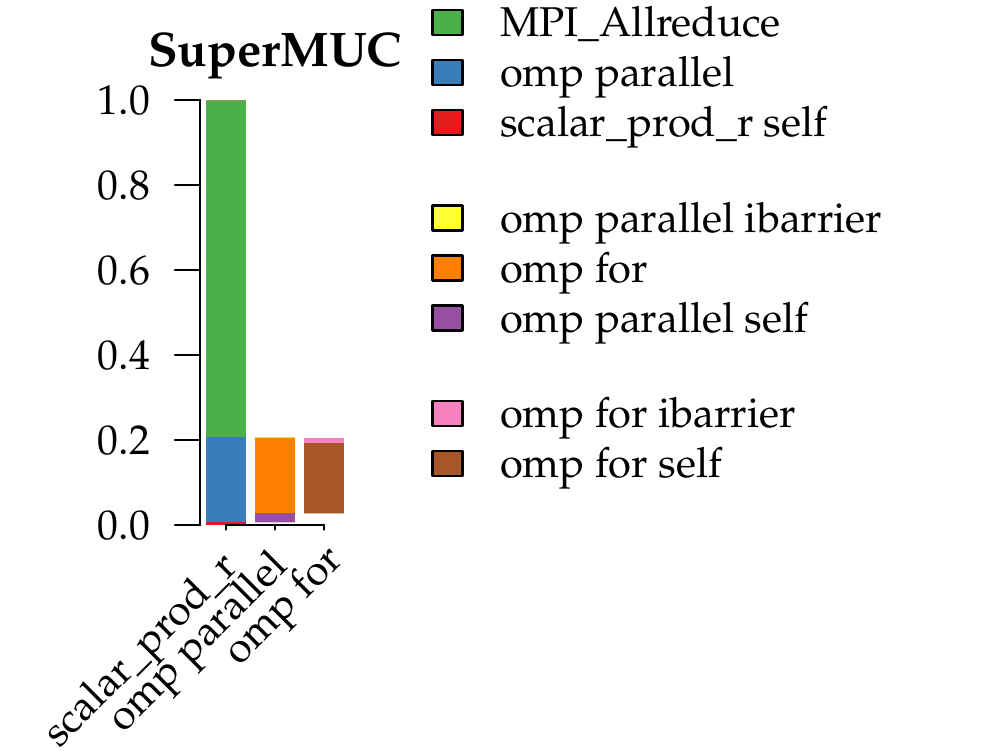}
	\end{subfigure}
	\caption{SCALASCA sampling of the sclar\_prod\_r function on 512 (256) nodes of BG/Q (SuperMUC), using 1 (2) tasks per node and 64 (8) threads per process during an inversion on a lattice volume of $48^3\times96$.}
	\label{fig:scalar_prod_r:scalasca}
\end{wrapfigure}

Finally, it is worthwhile to look at overheads on different machines and for different types of functions.
A SCALASCA sampling of the (real) scalar product of two spinor vectors is shown in figure \ref{fig:scalar_prod_r:scalasca}.
This computationally simple function suffers from substantial thread management overhead and a significant amount of time is spent in the barrier at the end of the for loop.
Whether optimization would result in measurable improvement seems to depend strongly on the architecture and number of threads, as shown by the comparison between BG/Q and SuperMUC, where the latter is dominated completely by the collective MPI communication.
More fine-grained measurements of these and other overheads are planned through SCALASCA event tracing.

\section*{Acknowledgements}

B.K. acknowledges full financial support by the National Research Fund, Luxembourg under AFR Ph.D. grant 27773315.
This work is supported in part by DFG and NSFC (CRC 110). This talk was part of a coding session sponsored partially by the
PRACE-2IP project, grant: RI-283493.

\bibliography{bibliography}{}
\bibliographystyle{PoS2}

\end{document}